# High Burst Rate Charging System for the Lithium Lens Power Supply

H. Pfeffer, D. Frolov, C. C. Jensen, M. E. Kufer, K. Quinn Jr
Fermilab
Batavia IL 60515
ccjensen@fnal.gov

*Abstract*—Two pulsed power systems have been upgraded for the g-2 experiment at Fermilab. The Pbar Lithium Lens supply previously ran with a half sine pulsed current of 75 kA peak, 400 us duration and a repetition rate of 0.45 pps. For the g-2 experiment, the peak current was reduced to 25 kA, but the repetition rate was increased to an average of 12 pps. Furthermore, the pulses come in a burst of 8 with 10 ms between each of 8 pulses and then a delay until the next burst. The charging rate has gone up by a factor of 20 due to the burst speed. A major challenge for the upgrade was to charge the capacitor bank while keeping the power line loading and charging supply cost to a reasonable level. This paper will discuss how those issues were solved and results from the operational system

*Keywords—pulse power, charging, burst, capacitor, resonant*

## I. Introduction

Beam operation to the Muon g-2 experiment started at Fermilab in spring 2017. The Fermilab accelerator complex is supporting neutrino physics with the NOvA and MicroBooNE experiments and the available time for beam for other experiments is limited. Improvements to the Proton Source now provide protons at up to 15 Hz, meeting the needs for both Neutrino experiments and g-2. A burst of 8 pulses of protons over 80 ms is followed by a gap of between 120 ms and 1 sec to the next burst of 8 pulses of protons to satisfy all experimental needs. The Muon Target Station has a lithium lens to focus the secondary particle beam and a pulsed dipole magnet for momentum-selecting the pions, which decay into the muons the experiment needs. Both the lens and pulsed magnet power supplies have been half sine wave pulsers since their original use in the old pbar complex [1,2]. However, new charging supplies were needed to meet the requirements for g-2 operation.

Taking a standard approach to recharging the pulse capacitor banks requires a very large capacitor charging supply. Even with charge recovery, the peak power required to recharge the lens capacitor bank in less than 10 ms is 120 kJ/s. This would also take a large peak power from the AC mains. Several charging schemes were investigated to meet the regulation requirement of +/- 0.1% and the charging time. Resonant charging by itself does not reduce the peak charging power since in a typical resonant charging scheme the bulk capacitor is recharged between pulses. Alternatively, an extremely large bulk capacitor, in size and cost, could be used so the the voltage does not droop substantially during the burst. We had faced similar challenges with a long pulse modulator where the solution was to let the capacitor bank droop substantially [3]. In this case we needed to solve the charging problem instead of the discharge problem.

## II. Resonant Charging

The bulk capacitor voltage is usually regulated to a precision of about 1% in a typical resonant charging system, Fig 1. Since the charge switch and D2 are typically combined into a thyristor this means that the pulse capacitor voltage will also be regulated to that level; the final pulse capacitor voltage is set by component values and initial conditions [4].

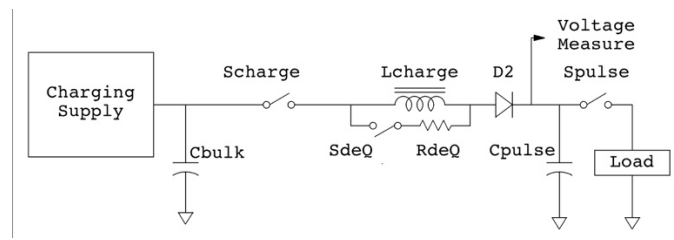

Fig. 1. Typical Resonant Charging Circuit

To achieve precision regulation with a standard resonant charging system, some excess energy is stored in the bulk capacitor and then dissipated with some form of deQ circuit, shown in Fig. 1 as SdeQ and RdeQ. The bulk capacitor voltage is maintained at a voltage that would lead to an over-charge of the pulse capacitor of several percent [5]. Then when the measured pulse capacitor voltage is at the desired level the deQ switch is triggered. This diverts the charging current into the deQ resistor and terminates the charging.

The energy remaining in the charging choke dissipates in the deQ resistor. The energy dissipated in diverting an extra 2% of the charge voltage is 4% of the energy supplied. This is the cost of precision regulation.

In pulsers with a constant repetition rate, the bulk capacitor is typically ten times the capacitance of the pulse capacitor. The bulk capacitor voltage declines by 10% while the pulse capacitor is being charged and is recharged by the charging supply before the next pulse. The charging rate is roughly constant so the power drawn from the AC mains is also roughly constant. The next charging pulse is at a known time and the charging supply can be sized for the average power with this in mind. The key requirement is the constant repetition rate.

Our application requires a burst of 8 pulses followed by a delay at least 12 times longer than the time between bursts. A typical solution could be a capacitor charging supply that directly charges the pulse capacitor without even using a bulk capacitor. We didn't want to use a charging supply capable of recharging either the bulk capacitor or the pulse capacitor in 10 ms because the peak power and cost are much larger than needed for the average power.

Let us first propose using a resonant charging system with a charging supply current and bulk capacitance sized so that the bulk capacitor only returns to its starting voltage prior to the next burst ( 200 ms after the first pulse). Now the bulk capacitor voltage continues to discharge with each pulse of the burst and regulation issues can occur. For example, chose a bulk capacitor 34 times larger than the pulse capacitor. It will discharge by 10% just before the final pulse of the burst, the same amount as in standard resonant charging. If we set this final voltage 2% above voltage margin for deQ, then for the first pulse of the burst it would be 12% high (10% + 2%) and the deQ resistor would have to absorb 24% of the charging energy. This may be a reasonable tradeoff in a low power system. The fundamental problem with the standard resonant charging circuits is that some of the energy stored in the charge inductor must be dissipated.

### III. LOSSLESS RESONANT CHARGING

The solution is to have no energy stored in the inductor when the pulse capacitor is at the final voltage. We call this "predictive deQ". A diode D1 and a current monitor on the charging choke are added as shown in Fig 2. The charging switch turns on as in standard resonant charging, but then it opens when a real-time computation indicates that the energy remaining in the charging choke is just enough to bring the partially charged pulse capacitor bank to the desired level, Eq. 1. The current in the charging choke continue to flow through D1 until D2 blocks and charging of the pulse capacitor is completed.

$$\tfrac{1}{2} C_{pulse} [V(t)_{pulse}]^2 + \tfrac{1}{2} L_{charge} [I(t)_{charge}]^2 = \tfrac{1}{2} C_{pulse} V_{final}^2 \quad (1)$$

The situation is now much different. There is now a lower limit to the ratio of bulk capacitor voltage to pulse capacitor voltage to complete charging, but within this range the pulse capacitor is charged to the required voltage without dissipating power in an additional element. The power supply can be sized to recharge the bulk capacitor just before the burst starts again by increasing the bulk capacitance. The charge inductor is basically the same as used in standard resonant charging.

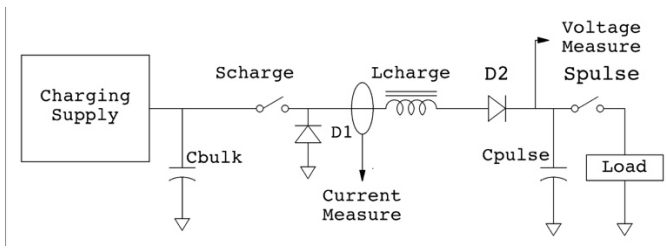

Fig. 2. Lossless Resonant Charging Circuit

### IV. DOUBLE DEQ

There are some practical limits to the regulation achievable with this technique. Saturation of the charging choke at higher currents changes the stored energy and requires knowledge of the choke saturation curve. When operating over a wide range of operating voltages on the pulse capacitor, the squaring function requires good dynamic range. Non-linearities in the measurement instrumentation or calculation can cause small regulation errors. These can be overcome by adding a more conventional deQ function to the circuit as shown in Fig 3. The component values shown are the ones used in the Litium Lens power supply.

As in standard deQ, we aim at a voltage 2% above the desired level and then trigger a lossy deQ circuit that regulates the final voltage precisely and dissipates the usual 4% loss. We call this sequence "Double deQ". Fig. 4 shows that the pulse current regulation achieved is within +/- 0.15%. The bulk capacitor voltage can be seen slowing decreasing during the burst and the charging choke current shape changes due to the different initial conditions for each pulse.

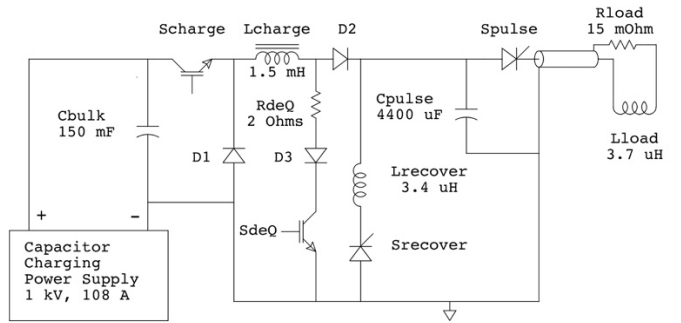

Fig. 3. Double DeQ Circuit with Design Values

There are some other features shown in Fig 3. The pulse capacitor reverses voltage during the circuit operation. We use charge recovery to bring the polarity of the pulse capacitor back to the normal polarity required for pulsing. The inductance of the recovery choke is about the same as the load because we have very little time to recover and charge. We use a thyristor based recovery circuit instead of a simple diode because the recover choke inductance is so small and the regulation requirements are so tight. The use of charge recovery also means we need to use D3 to block the diode built into the deQ switch. Minimal current flows in D1 during the recovery time because of the charging choke impedance.

### V. DESIGN TRADEOFFS

Choosing the inductance of the charging choke is based on the burst charging time and the minimum voltage of the bulk capacitor. The choke was specified and purchased in industry. The specification called out a smaller change in inductance at high current than might be typical of a filter choke and lower losses. The choke performed well even though the measured loss and saturation change were more than specified.

The amount of energy stored in the bulk capacitor and the size of the charging supply is a tradeoff in total size and cost. To minimize the total cost of the charging system, we assumed a cost of 1.0 $/W (peak) for the charging supply and 0.5 $/J for the bulk capacitor bank. Less expensive capacitor technology can be used for the bulk capacitor if the bank discharges less than 20% on a repetive bassis. The cost shown in Fig. 5 is for the charging system, power supply plus bulk capacitor bank, for various sizes of bulk capacitor bank for the burst rate conditions in our application. It does not include the cost of the pulse capacitor, recovery choke, the switches or the charging choke as those are independent of the charging scheme.

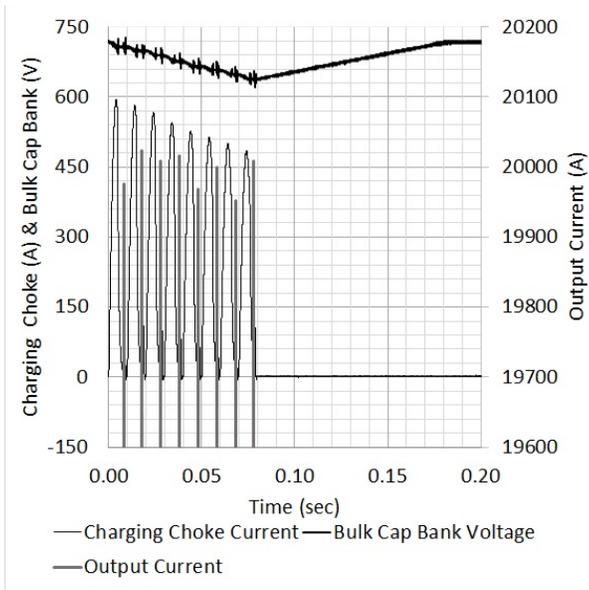

Fig. 4. Charging Current, Bulk Voltage and Output Current

The charging supply in this case is made of multiple units with the outputs in parallel. The supplies are in two groups on the secondaries of a Delta-Delta-Wye transformer to further reduce line harmonics. We chose a point very close to the minimum in cost based on the size of the charging supplies and the requirement to have an even number of them.

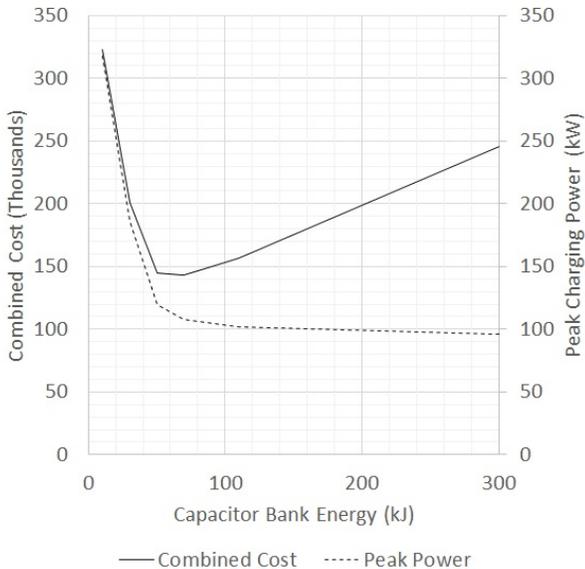

Fig. 5. Total Cost and Peak Charging Power vs Bulk Capacitor Stored Energy

## VI. Conclusions

A new charging topology has been described that reduces the peak power of a charging supply with a burst rated load by allowing the bulk voltage to vary during the burst. A related benefit of this topology is the ability to reduce the cost of the charging system. The scheme can also be used to allow for greater line variation with any repetition rate load. Finally, this method can allow for a tradeoff between increasing efficiency and an inductor with less saturation.

The Lithium Lens power supply, and a similiar supply for the pulsed magnet, have been running for 6 months now. The only significant failure has been an instance where the charging switch and pulse switch turned on simultanesouly. A hardware exclusive OR has been added at the switch driver output to inhibit that failure. There have also been some small changes made to the circuit to improve performance when re-starting operation after a pause.

## Acknowledgment

We would like to thank Dan Wolff for his contributions and comments during the design of the circuit, Kevin Martin for implementing computer controls and Jim Biggs, Bryan Falconer, Nick Gurley, Mike Henry, and Kevin Roon for their contributions, suggestion and hard work during the construction and testing of the power supply.